\documentclass[prl,aps,twocolumn,english]{revtex4}
\usepackage{graphicx}
\usepackage{amssymb, amsmath, color, rotating}

\begin{document}

\title{Anomalous Spin segregation in a weakly interacting  two-component  Fermi gas}

\author{Stefan S. Natu}
\email{ssn8@cornell.edu}
\author{Erich J. Mueller}
\email{em256@cornell.edu}
\affiliation{Laboratory of Atomic and Solid State Physics, Cornell University, Ithaca, New York 14853, USA.}

\begin{abstract}
We explain the spin segregation seen at Duke in a two-component gas of $^6$Li \cite{jt} as a mean-field effect describable via a collisionless Boltzmann equation. As seen in experiments, we find that slight differences in the trapping potentials in the two spin states drive small spin currents. Hartree-Fock type interactions  convert these currents into a redistribution of populations in energy space, and consequently a long lived spin texture develops. We explore the interaction strength dependence of these dynamics, finding  nontrivial dependence on system parameters and close agreement with experiment. 
 \end{abstract}
\maketitle

Spin waves in dilute cold gases such as spin polarized Hydrogen ~\cite{johnson, levy} or $^{87}$Rb \cite{lew, oktel, clark, laloe2} are driven by exchange effects and are manifestations of quantum coherence in a non-degenerate gas. Recent experiments at Duke ~\cite{jt} on $^{6}$Li, with the interactions tuned to be very small, saw an unexpected spin segregation describable as a very long wavelength, low frequency, longitudinal spin wave. Their attempts to explain this behavior in terms of hydrodynamic spin-wave theory failed. Here we use a collisionless Boltzmann equation  to explain this anomalous phenomenon. Our approach was motivated by discussions with John Thomas, who has recently explored a simplified version of this theory in a work with Du, Luo, and Clancy \cite{jtnew}.  A concurrent study by Pi\'echon, Fuchs, and Lalo\"e \cite{laloe} reached similar conclusions. Spin waves in this collisionless Knudsen regime have been studied extensively, both theoretically and experimentally \cite{bigelow, silin,leggett, laloeorig, laloeorig2}. 

In the Duke experiments \cite{jt}  a cold gas ($T/T_{F} \sim 4$)  of roughly $2 \times 10^{5}$ $^6$Li atoms, in their lowest hyperfine state (denoted $\sigma = \downarrow$), were prepared in an optical plus magnetic trap with a trapping potential of the form $U_{\downarrow} = \frac{1}{2}m\omega_{R}^{2}r^{2} +\frac{1}{2}m\omega_{Z, \downarrow}^{2}Z^{2}$, with $\omega_{R}  =2 \pi \times 4360$Hz and  $\omega_{Z,\downarrow} = 2\pi \times 145$Hz. A radio pulse was used to coherently transfer atoms into a superposition of the $\downarrow$ and the next hyperfine level (denoted $\uparrow$). The subsequent dynamics were observed for several different bias magnetic fields, hence several different scattering lengths. 

When the scattering length was small and negative they observed that after $\sim 100$ms of evolution, the two components of the gas segregate axially with the $\uparrow$ component moving inward, and the other moving outward. This spin texture persisted on timescales of a few seconds, much longer than the timescale for small oscillations. When the sign of the scattering length was changed, the  $\uparrow$ moved outward and the $\downarrow$ moved inward. When the scattering length was tuned to zero, no dynamics was seen. 

\begin{figure*}[htbp]
\begin{picture}(250, 150)(10, 10)
\put(-45, 165){\textbf{(a) $\bf{a}~\bf{>}~\bf{0}$}}
\put(115, 165){\textbf{(b) $\bf{a}~\bf{=}~\bf{0}$}}
\put(285, 165){\textbf{(c) $\bf{a}~\bf{<}~\bf{0}$}}
\put(125, 5){\Large{\textbf{$Z/\delta Z$}}}
\put(-115, 50){\Large{\begin{sideways}\textbf{$s(R, t)/s(0, 0)$}\end{sideways}}}
\put(-100, 20){\includegraphics[scale=0.5]{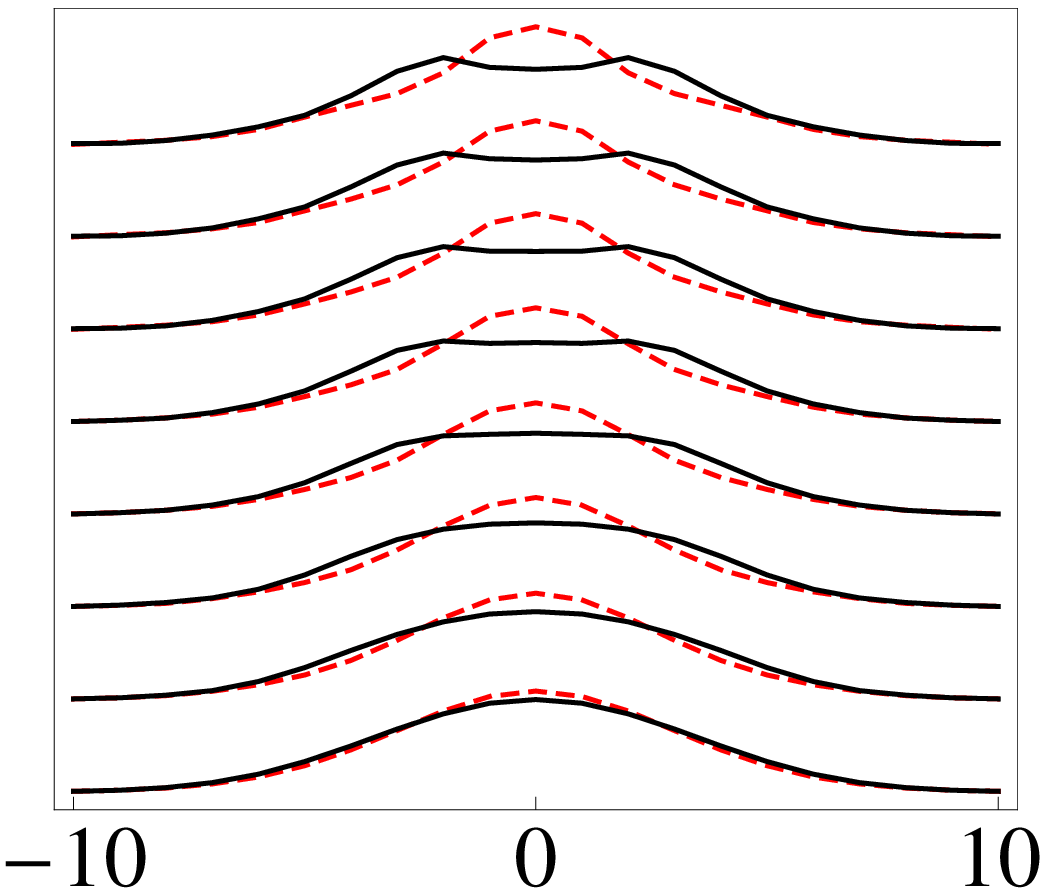}}
\put(65, 20){\includegraphics[scale=0.5]{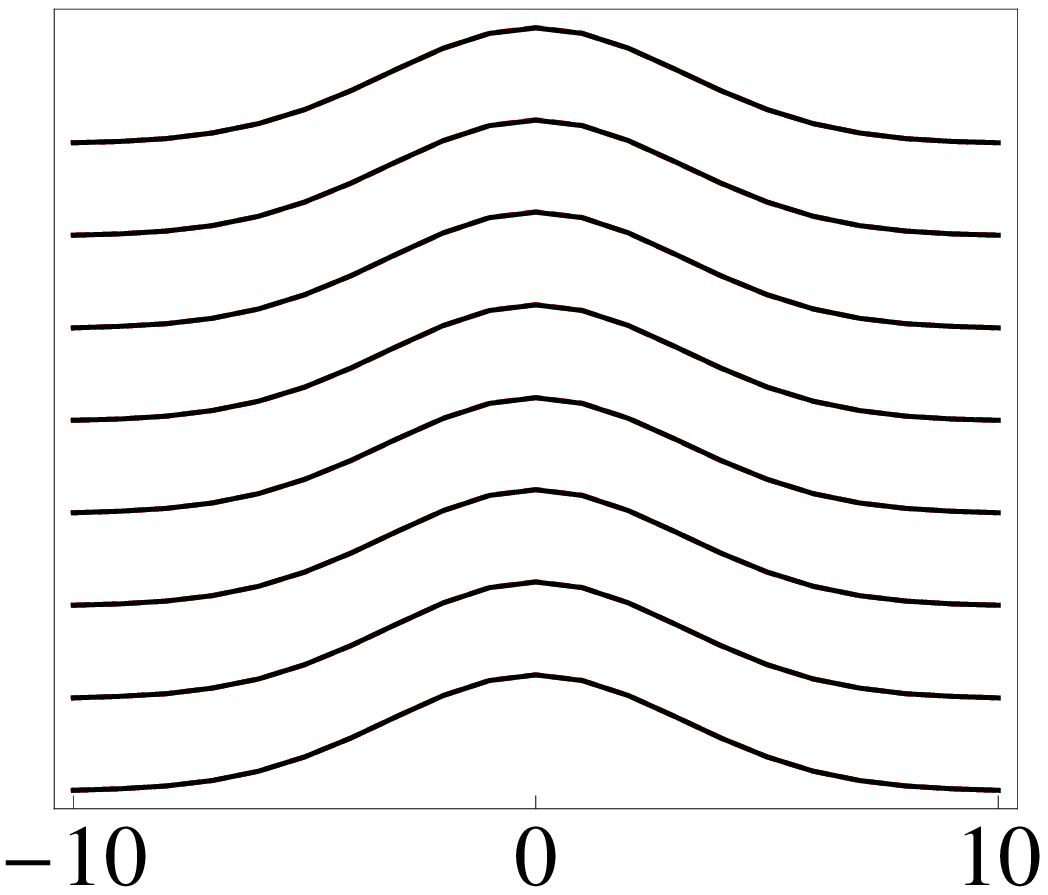}}
\put(235, 20){\includegraphics[scale=0.5]{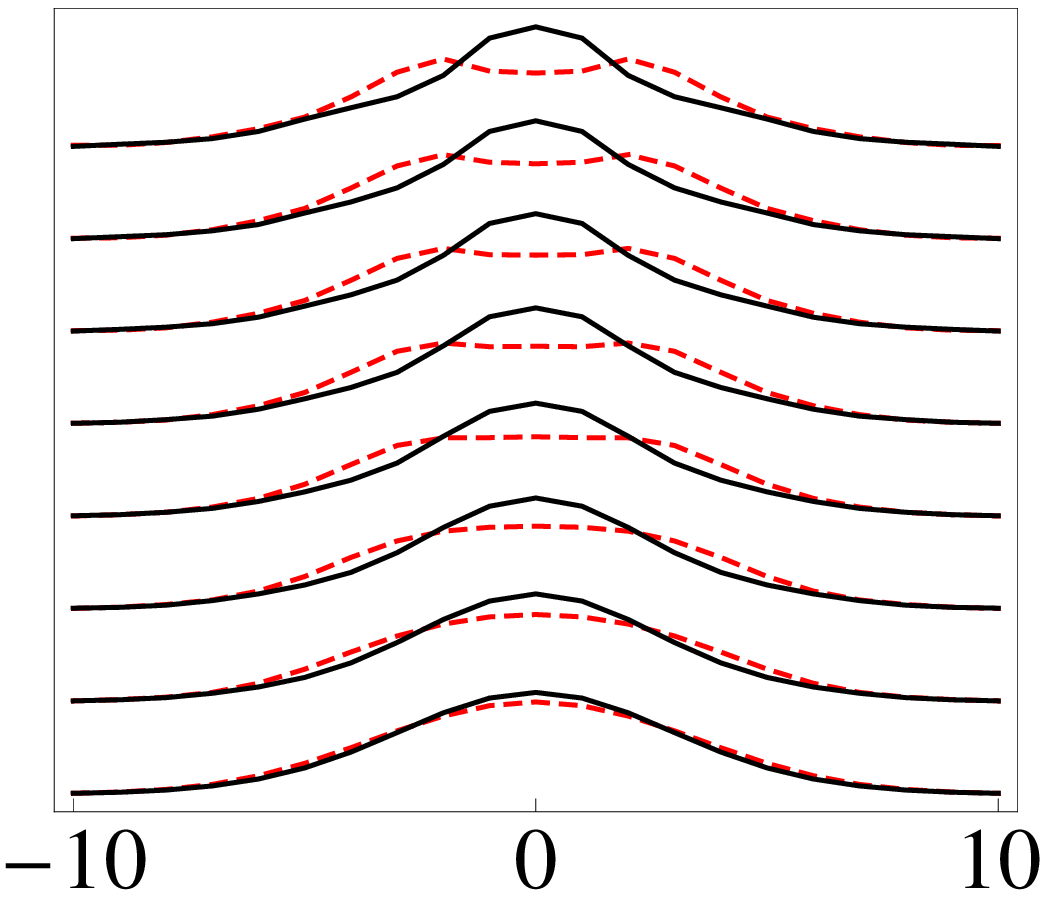}}
\end{picture}
\caption{\label{fig:-1} (Color Online) Time evolution of one-dimensional densities $s_\uparrow$ and $s_\downarrow$: The up- (black, solid), and down- (red, dashed) spin density in space (horizontal axis in units of $ \delta Z =20~(m \omega_{z})^{-1/2}$) for (a) $a = 4.5a_{B}$, (b) $a = 0$ and (c) $a = -4.5a_{B}$, where $a_{B}$ is the Bohr radius. Time runs from bottom to top, with each offset density profile is separated by $25$ms.  No dynamics are seen on this scale for $a=0$.}
\end{figure*}

This behavior was surprising. According to hydrodynamic spin wave theory \cite{levy, oktel}, the characteristic timescale for any oscillations should be given by the axial oscillator time, roughly two orders of magnitude faster than the observed dynamics. Moreover the only possible mechanism for driving this spin segregation is the very slight difference in the axial trap potential $\frac{d^{2}}{dz^2}[(U_{\uparrow} - U_{\downarrow})] \sim 2\pi (4.4 \times 10^{-4})$Hz/$\mu$m$^{2}$. It is surprising that such a small difference in the trap leads to such dramatic density redistributions. 

An intuitive picture of these dynamics is presented by Du \textit{et al.} \cite{jtnew}.  They note that since the time-scale of spin rearrangement is long compared to the oscillation period, local equilibrium is never attained. Instead, each atom's spin dynamics is controlled by a mean field, averaged over its periodic trajectory. Low energy atoms that spend more time in high density regions experience a greater mean field. The $\downarrow$ atoms see a slightly weaker trapping potential and hence, for a given energy, have trajectories which extend over more space. This results in those atoms seeing a smaller mean field. The net result of the subsequent dynamics is a spin segregation in energy space. 

Here we show how this behavior plays out in phase space. Following standard arguments \cite{silin, jeon},  we derive an effective  1D collisionless Boltzmann equation. Solving this equation numerically, we reproduce the experimental observations.

At the temperatures of interest ($T \sim 27\mu$K) one only needs to consider s-wave scattering and the
Hamiltonian reduces to
\begin{widetext}
\begin{equation} \label{eq:1}
\hat{H}(t) = \sum_{\sigma = \uparrow, \downarrow} \int d \textbf{r}~\Psi^{\dagger}_{\sigma}(\textbf{r}, t) (- \frac{1}{2m}\nabla_{r}^{2} + U_{\sigma}(r))\Psi_{\sigma}(\textbf{r}, t)  + g\int d \textbf{r}_{1}\Psi^{\dagger}_{\uparrow}(\textbf{r}_{1}, t)\Psi^{\dagger}_{\downarrow}(\textbf{r}_{1}, t)\Psi_{\downarrow}(\textbf{r}_{1}, t)\Psi_{\uparrow}(\textbf{r}_{1}, t),
\end{equation}
\end{widetext}
where the field operators obey the fermionic equal time anti-commutation relations $\{ \Psi^{\dagger}_{\sigma}(\textbf{r}_{1}, t),\Psi_{\sigma^{'}}(\textbf{r}_{2}, t)\} = \delta(\textbf{r}_{1} - \textbf{r}_{2})\delta_{\sigma, \sigma^{'}}$, and $g = \frac{4 \pi a}{m}$ with s-wave scattering length $a$. Near the magnetic fields of interest $a(B) = -3.5(B-B_{0})a_{B}/G$ ~\cite{jt}.  We have set $\hbar = 1$ throughout, and we work in the Larmor frame rotating with a frequency equal to that of the $\downarrow \rightarrow\uparrow$ transition for a uniform gas.

Given the small scattering lengths, and low densities in this experiment ($a \sim 4a_{B}$, $n \sim 10^{12}$cm$^{-3}$), the mean collision time $\tau = 1/(n \sigma v) \sim 10$ s is much longer than the timescale of the experiment, and interactions only enter at the mean field level. From experimental studies of relaxation in a single component gas ~\cite{jt}, it appears that the
time between background collisions $\tau_b$, due to an imperfect vacuum, is  also on the order of several seconds. For times short compared to $\tau$ and $\tau_b$, one can describe the system in terms of  a collisionless Boltzmann equation, where the system is described by  a Hartree-Fock approximation. The long timescales involved in collisional relaxation also explain why a simple hydrodynamic theory does not capture the physics of the phenomenon. 
\begin{figure}[hbtp]
\begin{picture}(150, 250)(10, 10)
\put(-35, 265){\textbf{(a)}}
\put(-35, 125){\textbf{(b)}} 
\put(0, 155){\includegraphics[scale=0.65]{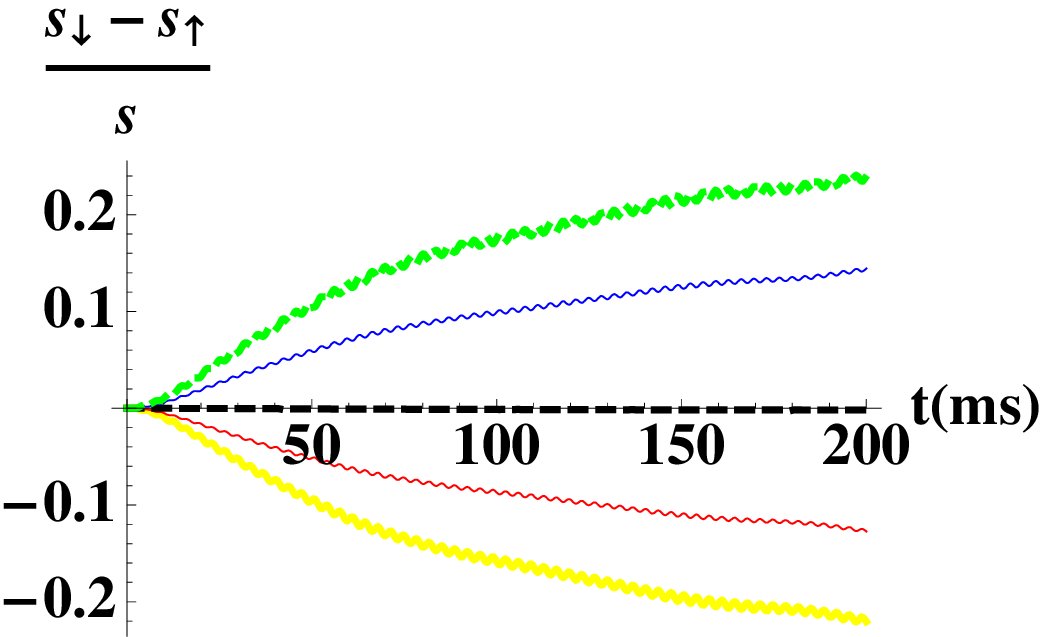}}
\put(0, 5){\includegraphics[scale = 0.65]{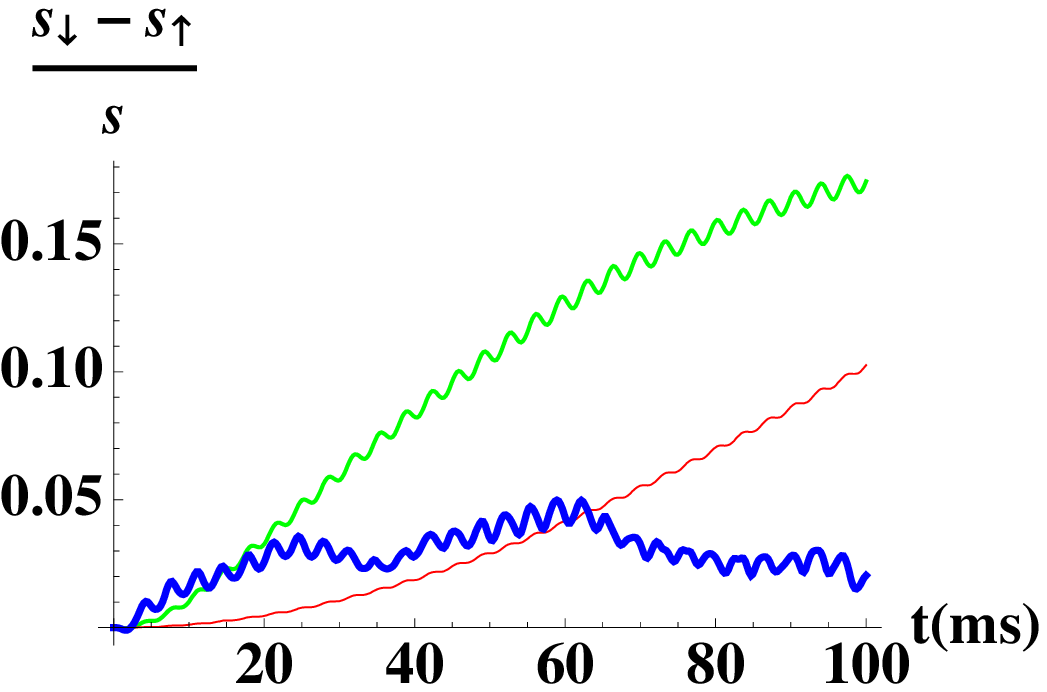}}
\end{picture}
\caption{\label{fig:-2} (Color Online) Top: Central spin density difference ($s_z(0, t)=s_\downarrow(0, t)-s_\uparrow(0, t)$) normalized to the total central density ($s= s_{\uparrow}(0) + s_{\downarrow}(0)$) for various scattering lengths. The total central density is constant in time. Bottom to top:  $a = -4.55~a_{B}$, $a = -2.45~a_{B}$, $a = 0~a_{B}$, $a = 2.8~a_{B}$, and $a = 4.55~a_{B}$. (b): First $100$ms of evolution of $s_{z}/s$ for $a =4.55~a_{B}$, and different values of $\delta \omega  = \omega_{\uparrow}-\omega_{\downarrow}$: thin (red) -- $2\pi \times 0.15$~mHz,  green -- $2\pi \times 1.5$~mHz, thick (blue) -- $2\pi \times 15$~mHz. The green curve corresponds to the experimental value of $\delta \omega$. }\end{figure}

\begin{figure}
\begin{picture}(280, 280)(10, 10)
\put(50, 270){{\textbf{$f_{\uparrow\uparrow}+f_{\downarrow\downarrow}$}}}
\put(135, 270){\textbf{$f_{\uparrow\uparrow}$}}
\put(210, 270){\textbf{$f_{\downarrow\downarrow}$}}
\put(25, 35){\includegraphics[scale = 0.25]{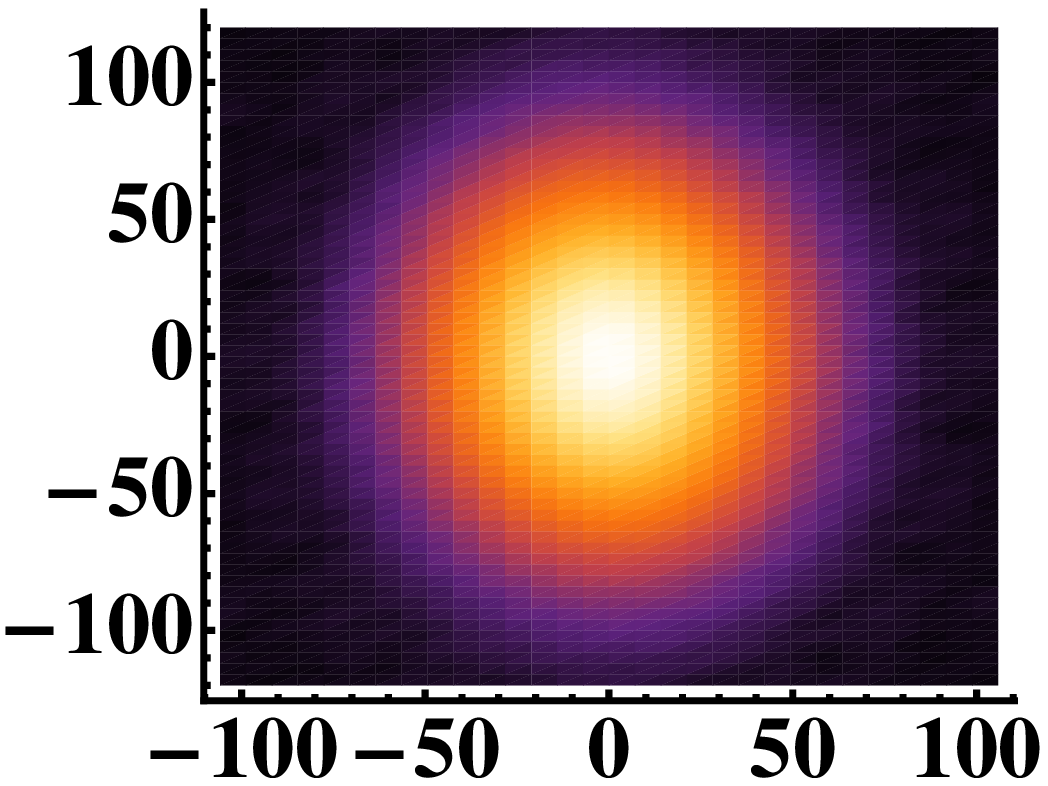}}
\put(25, 115){\includegraphics[scale = 0.255]{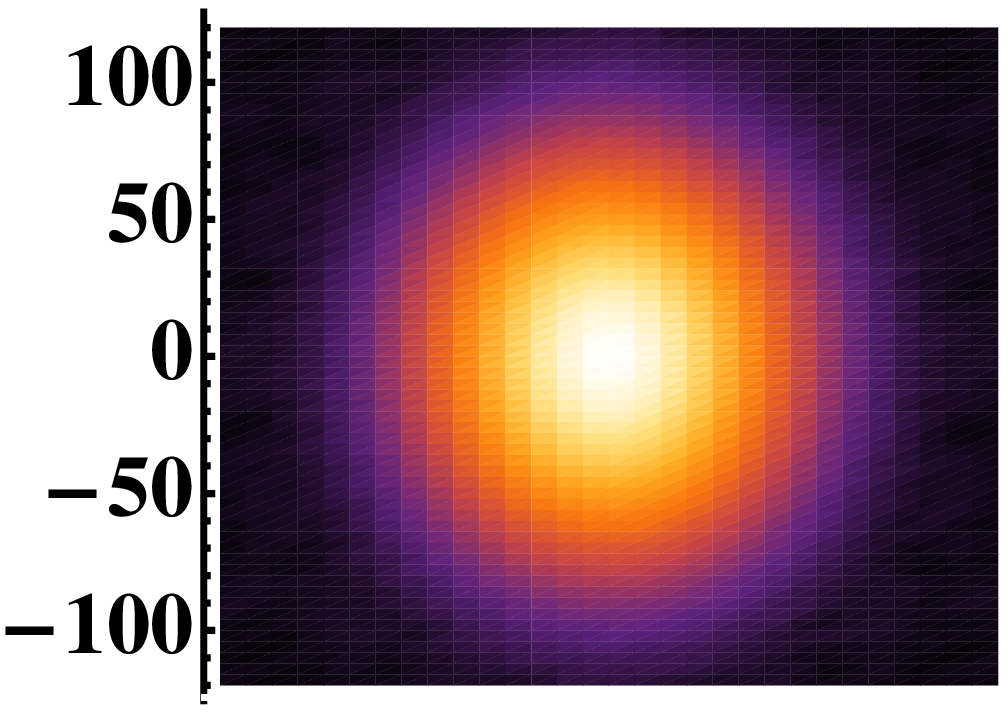}}
\put(25, 195){\includegraphics[scale = 0.255]{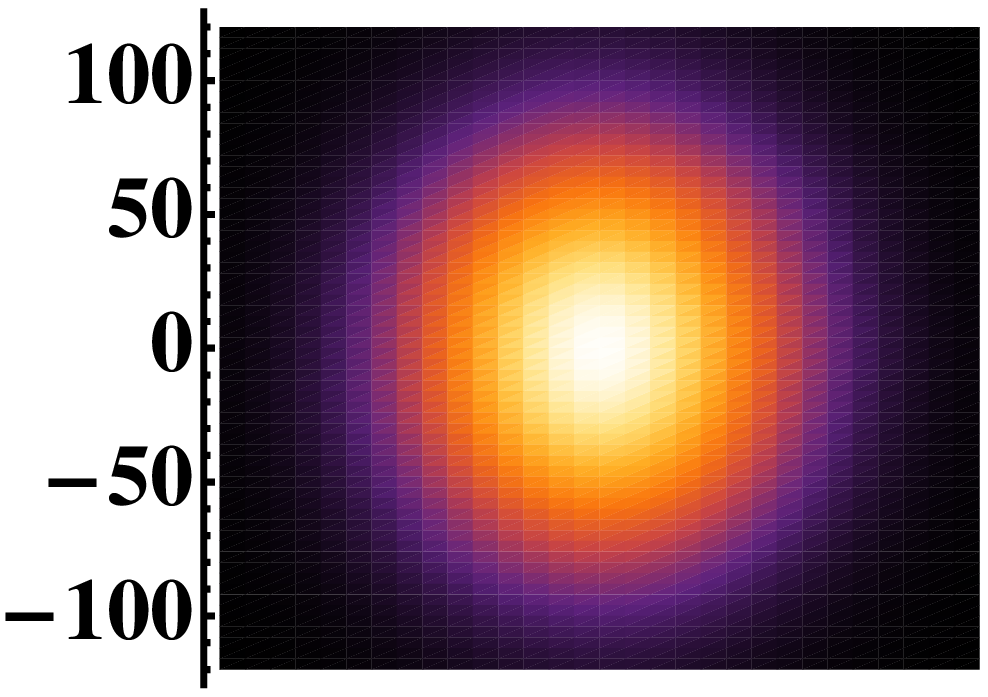}}
\put(100, 35){\includegraphics[scale = 0.25]{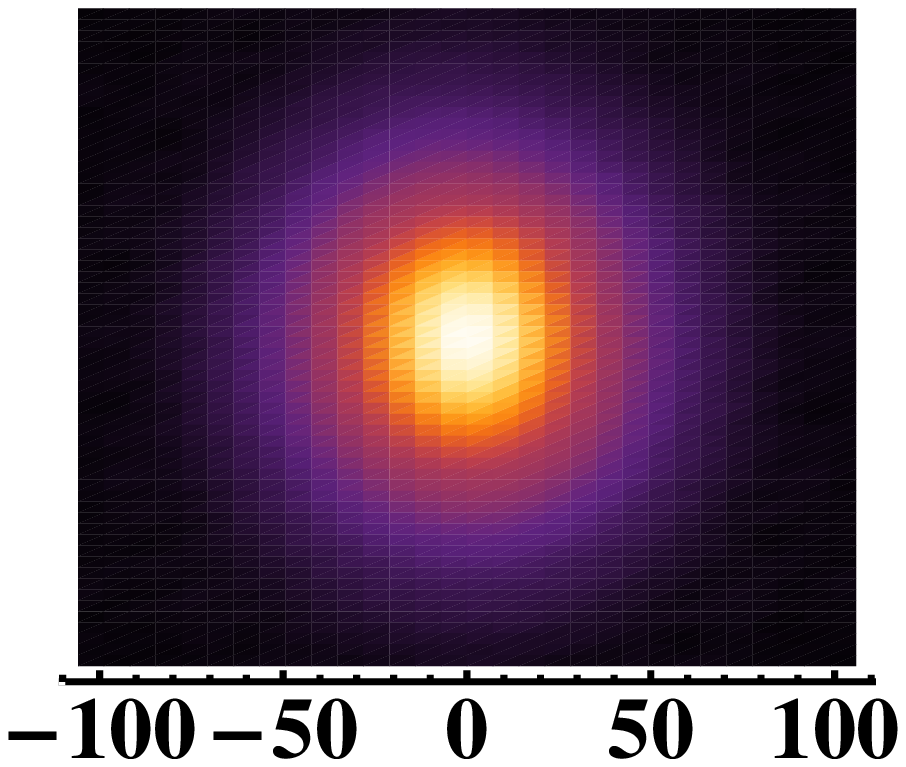}}
\put(100, 114){\includegraphics[scale = 0.26]{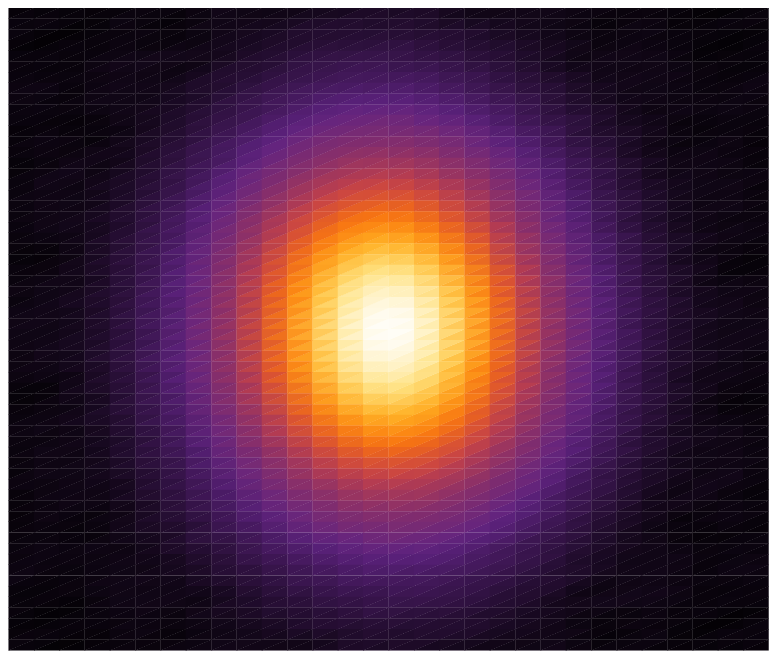}}
\put(100, 192){\includegraphics[scale = 0.26]{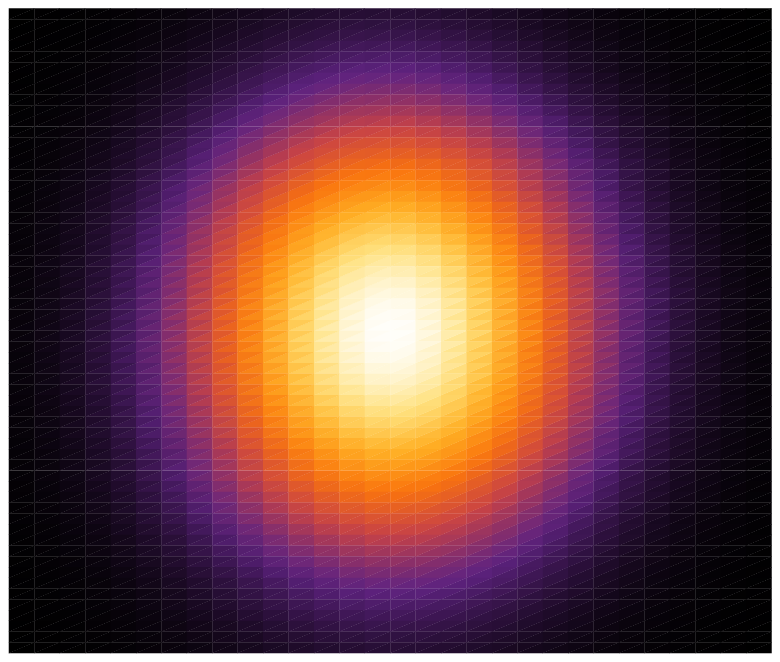}}
\put(175, 35){\includegraphics[scale = 0.25]{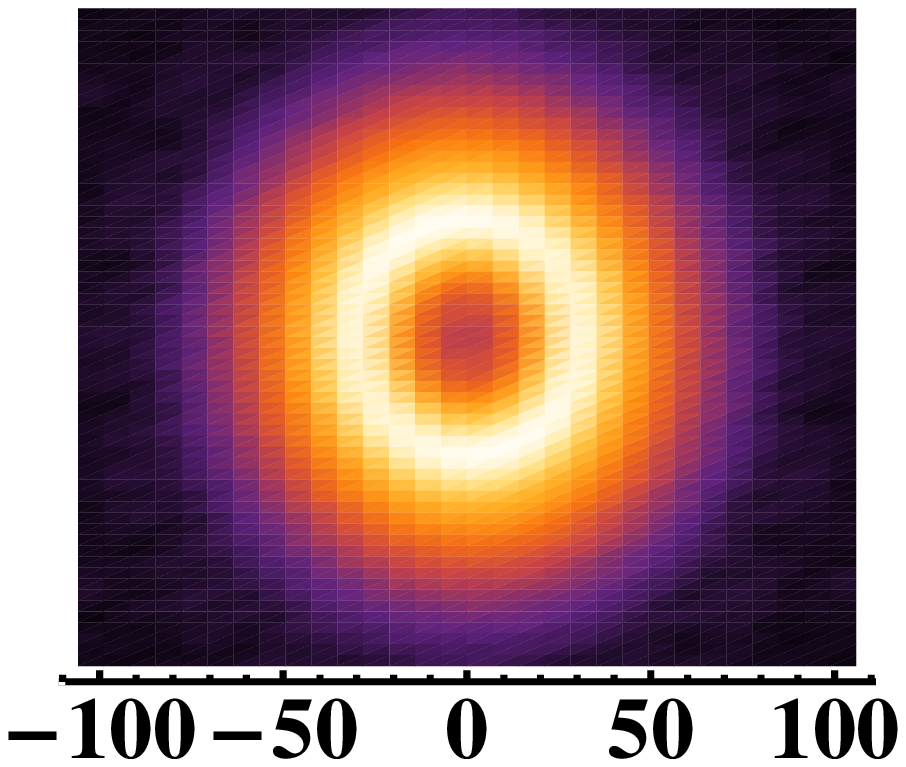}}
\put(175, 114){\includegraphics[scale = 0.26]{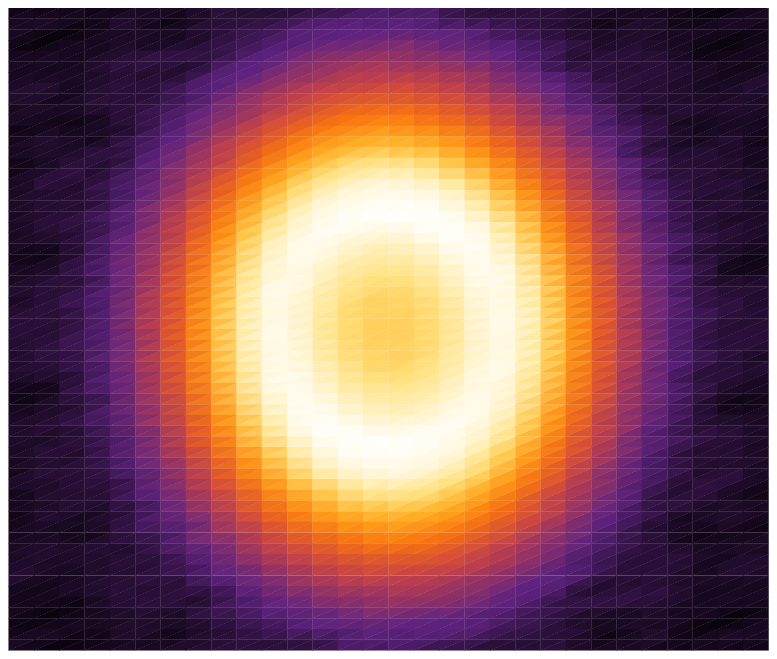}}
\put(175, 193){\includegraphics[scale = 0.26]{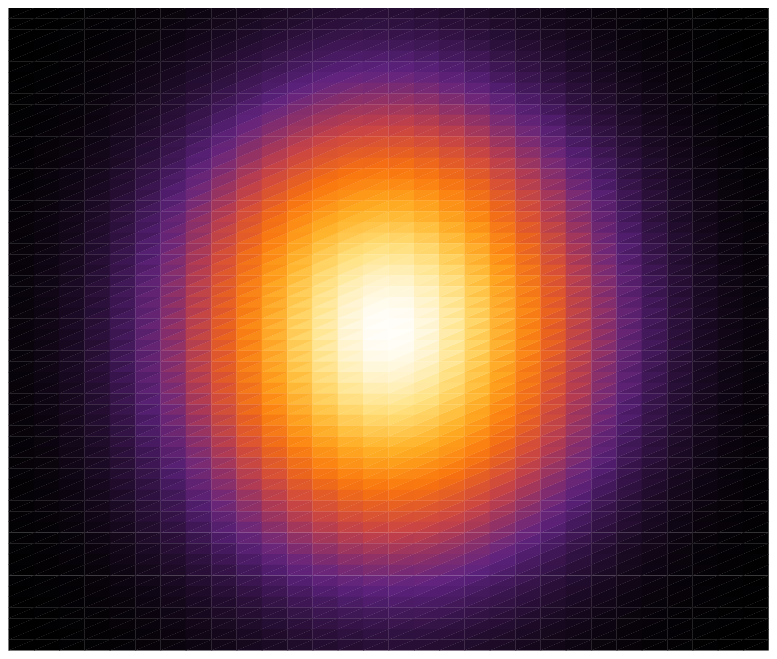}}
\put(115, 15){\Large\textbf{$Z\sqrt{m \omega_{z}}$}}
\put(08, 125){\Large\begin{sideways}\textbf{$p_{Z}/\sqrt{m \omega_{z}}$}\end{sideways}}
\end{picture}
\caption{\label{fig:-3} (Color Online) Spin segregation in phase space. Lighter colors represent higher density. The total phase space distribution (left column), up-spin distribution (middle column) and down spin distribution (right column) at $t =0$ (top row), $t = 100$ms (center row) and $t = 200$ms (bottom row) for $a = -4.5~a_{B}$ are shown. The phase space distribution is only a function of energy, but it is not of a simple Boltzmann form.}
\end{figure}

Following \cite{kadbaym} we use the Heisenberg equations for $\Psi_{\sigma}$(r, t) to derive the equations of motion for the
spin dependent Wigner function
\begin{eqnarray}\label{eq:2}
\overleftrightarrow{\textbf{F}} &=& 
\left(\begin{array}{cc} 
f_{\uparrow\uparrow}(\textbf{p}, \textbf{R}, t)&f_{\uparrow\downarrow}(\textbf{p}, \textbf{R}, t) \\ 
f_{\downarrow\uparrow}(\textbf{p}, \textbf{R}, t)&f_{\downarrow\downarrow}(\textbf{p}, \textbf{R}, t) \end{array}\right)\\
\nonumber
f_{\sigma \sigma^{'}}(\textbf{p}, \textbf{R}, t) &=& \int d \textbf{r} e^{-\imath \textbf{p}\cdotp \textbf{r}} \langle \Psi^{\dagger}_{\sigma}(\textbf{R} - \frac{\textbf{r}}{2}, t) \Psi_{\sigma^{'}}(\textbf{R} + \frac{\textbf{r}}{2}, t) \rangle, 
\end{eqnarray}
which is the quantum analogue of the classical distribution function. Here $\textbf{p}$ represents the momentum, $\textbf{r} = \textbf{r}_{1} - \textbf{r}_{2}$ is the relative coordinate and $\textbf{R} = \frac{\textbf{r}_{1}+\textbf{r}_{2}}{2}$ is the center of mass coordinate.  Treating interactions in the Hartree-Fock approximation \cite{kadbaym}, the resulting equations involve densities and currents such as
$\langle s_{\sigma, \sigma^{'}}(\textbf{R}, t)\rangle = \langle \Psi^{\dagger}_{\sigma}(\textbf{R}, t) \Psi_{\sigma^{'}}(\textbf{R},  t) \rangle = \int \frac{d \textbf{p}}{(2\pi)^{3}}f_{\sigma\sigma^{'}}(\textbf{p}, \textbf{R}, t)$ and  $\langle\textbf{j}_{\sigma\sigma^{'}}(\textbf{R}, t) \rangle  = \int \frac{d\textbf{p}}{(2\pi)^{3}}\textbf{p} f_{\sigma\sigma^{'}}(\textbf{p}, \textbf{R}, t)$. 

We define $s_{\sigma\sigma} = s_{\sigma}$, $s_{\uparrow\downarrow} = s_{+}$, $s_{\downarrow\uparrow} = s_{-}$ and analogously for the spin currents. Here $s_{\pm}$ refer the spin raising and lowering operators that are related to the transverse components of the spin $s_{x}$ and $s_{y}$ in the usual way $s_{\pm} = s_{x} \pm \imath s_{y}$. Throughout this paper we use upper-case letters to denote the components of position and momentum, and lower-case  letters to denote the spin degrees of freedom.
These transverse spin components represent a quantum coherence which is not captured by a classical model of a two component gas.
The z-spin density is $s_{z}(\textbf{R}, t) = s_{\downarrow}(\textbf{R}, t) - s_{\uparrow}(\textbf{R}, t)$, and the particle number is $N = \int d\textbf{R}~s(\textbf{R}, t)$ where $s(\textbf{R}, t) = s_{\uparrow}(\textbf{R}, t) +s_{\downarrow}(\textbf{R}, t)$. 
Assuming that all potentials are slowly varying in space and time, one finds
\begin{equation}\label{eq:3}
\frac{\partial}{\partial t}\overleftrightarrow{\textbf{F}} + \frac{\textbf{p}}{m}\cdotp\nabla_{\textbf{R}}\overleftrightarrow{\textbf{F}} = \imath [\overleftrightarrow{\textbf{V}}, \overleftrightarrow{\textbf{F}}] +  \frac{1}{2}\{\nabla_{\textbf{R}}\overleftrightarrow{\textbf{V}},\nabla_{\textbf{p}}\overleftrightarrow{\textbf{F}}\}
\end{equation}
where the potential matrix is 
\begin{equation}\label{eq:4}
\overleftrightarrow{\textbf{V}} =  \left(\begin{array}{cc} U^{eff}_{\uparrow}&-g s_{+}\\ -g s_{-}&U^{eff}_{\downarrow}
\end{array}\right)
\end{equation}
and the effective potentials are $U^{eff}_{\uparrow}(R, t) = U_{\uparrow}(R) + g s_{\downarrow}(\textbf{R}, t)$ and $U^{eff}_{\downarrow}(R) = U_{\downarrow}(R) + g s_{\uparrow}(\textbf{R}, t)$. Commutators and anti-commutators are respectively given by square brackets and braces. The diagonal terms of the potential matrix include the direct contribution to forward scattering, while the off diagonal components represent the exchange contribution.

Given the differences between the radial and axial trapping frequencies, the relevant dynamics of the system are one-dimensional.  In the nondegenerate limit that we consider here one can reduce (Eq.~\ref{eq:4}) to a one dimensional Boltzmann equation by making the ansatz: $f(\textbf{p}, \textbf{R}, t) = f(p_{\perp}, p_{Z}, R_{\perp}, Z, t) = f(p_{\perp},R_{\perp}) f(p_{Z}, Z, t)$, where the notation $\perp$ has been used to denote the transverse directions. The initial distribution is assumed to be a stationary state of the non-interacting Boltzmann equation for spin-down particles:
\begin{equation}\label{eq:5}
f_{\sigma\sigma^{'}} = \frac{A}{2}e^{-\beta\left({p_{Z}^{2}}/{2m} + U_{\uparrow}(Z)\right)}e^{-\beta\left({p_{\perp}^{2}}/{2m} + U(R_{\perp})\right)}
\end{equation}
where the prefactor $A = N \omega_{z}\omega_{r}^{2}/(k_{B} T)^{3}$ is defined such that the constraint $N = \int d\textbf{R} (s_{\uparrow}(\textbf{R}, t) +s_{\downarrow}(\textbf{R}, t))$ is satisfied.

By assuming the above Gaussian density profile, we express the 3D density as $s^{3D}_{\sigma\sigma^{'}}(\textbf{p}, R_{\perp}, Z, t)= \frac{A}{2}\int \frac{d\textbf{p}}{(2 \pi)^{3}}f_{\sigma\sigma^{'}}(\textbf{p}_{\perp}, R)f_{\sigma\sigma^{'}}(p_{Z}, Z, t) = \int \frac{d^{2}p_{\perp}}{(2 \pi)^{2}}f_{\sigma\sigma^{'}}(p_{\perp}, R)s^{1D}_{\sigma\sigma^{'}}(Z, t)$, where $s^{1D}_{\sigma\sigma^{'}}(Z,t) = \frac{A}{2}\int\frac{d p_{Z}}{2 \pi}f_{\sigma\sigma^{'}}(p_{Z}, Z, t)$.~Spatial averaging in the transverse direction renormalizes the coupling constant by $\frac{1}{2}$ \cite{oktel}. Finally, integrating the transverse momentum coordinates yields an effective interaction strength $g^{1D} = \frac{1}{8\pi^{2}} g N (m\omega_{z})(\omega_{r}/k_{B} T)^{2}$.

\textit{Results and Discussion}:$-$~
Working in units where lengths and momenta are measured in terms of 
 the oscillator length $(m \omega_{z})^{-1/2}$ and $(m k_{B} T)^{1/2}$ respectively, we use a phase space
 conserving split-step approach \cite{teuk} to integrate
 Eq.~(\ref{eq:4}), using a $20$ by $20$ by $800$ grid in $Z$-$p_{Z}$-$t$ with step sizes $ \delta Z =20~(m \omega_{z})^{-1/2}$, $\delta p_{Z} = 0.6~(m k_{B} T)^{1/2}$, and $\delta t = 0.04\frac{2\pi}{\omega_{z}}$. We verified that our grid was sufficiently fine so that our results no longer depended on the step sizes.
  
In Fig.~\ref{fig:-1} we show the time evolution of the density profile for the up (black/solid curve) and down spins (red/dashed curve) in space for the first $200$~ms, initializing the cloud in an $s_{x}$ state. We plot the behavior for three different values of the scattering length, finding that both the magnitude and timescales are in agreement with experiment \cite{jt}.

In Fig.~\ref{fig:-2}(a), we quantify the magnitude of the spin segregation by plotting the central density difference ($s_z=s_\downarrow-s_\uparrow$), normalized to the total density ($s=s_\uparrow+s_\downarrow$) as a function of time for a range of scattering lengths. 
This quantity peaks near $t\sim 200$ ms.  We extract the timescale associated with spin segregation by taking the slope of the graph at small times (Fig.~\ref{fig:-2}(b) green curve) to find $ \frac{d}{d t} \frac{s_{z}(0, t)}{s} \sim 1/200$~(ms)$^{-1}$ for $a = 4.55 a_{B}$ . Furthermore, the figure reveals oscillations in the spin density difference at a frequency $\sim 2\pi 300$~Hz ($\sim 2 \omega_{z}$) that is weakly dependent on the interaction strength, corresponding to the lowest breathing mode of a two component Fermi gas \cite{pethicksmith}. The amplitude of these oscillations depends on the difference in the trap frequencies seen by the $\uparrow$ and $\downarrow$ atoms.  These oscillations are not captured in the analysis presented in \cite{jtnew}. 

In previous experiments on bosons  \cite{lew}, the spin dynamics were much faster than such collective modes.  This difference can be attributed to the ratios of the mean field interaction energy to the trap frequency
$\lambda=g^{1D}/\omega_{z}$.  In the current experiment $\lambda\sim 0.2$ while in \cite{lew}, $\lambda\sim 10$.

Fig.~\ref{fig:-2}(b) shows that both the magnitude and timescale for spin segregation seen in \cite{jt} is strongly dependent on the difference in trapping frequencies ($\delta \omega = \omega_{\uparrow}-\omega_{\downarrow}$). Had this frequency difference in \cite{jt} been an order of magnitude larger (blue/ thick curve in Fig.~\ref{fig:-2}(b)), the dynamics would have been much more complicated and much less dramatic. 

As previously discussed, an important observation in \cite{jtnew} was that the
 spin segregation in \cite{jt} can be viewed as a segregation in energy space. We illustrate this effect in  Fig.~\ref{fig:-3} by plotting the phase space distributions for $a = -4.5a_{B}$ for $t = 0$, $100$ and $200$ ms respectively.  One sees that the phase space distributions are not separately functions of $Z$ and $p_Z$, but instead depend on $\omega_z^2 Z^2+p_Z^2/m$. 

Finally we note that this spin segregation is very robust. We can illustrate this by exciting a large amplitude spin dipole mode at $t=0$. We find that spin segregation occurs even as the $\uparrow$ and $\downarrow$ atoms slosh around in the trap, out of phase with one another. As may be expected for a gas in the Knudsen regime, oscillations on timescales much shorter than the interactions do not change the long term dynamics.  Nonetheless, it would be interesting to observe this stability experimentally. 
 
\textit{Summary and Conclusions}:$-$~
Using standard techniques \cite{kadbaym}, we have derived a collisionless Boltzmann equation which reproduces the anomalous spin waves seen in \cite{jt}.  This is an exciting regime for spin waves, as the system is far from local equilibrium.  Remarkably we find an ergodicity where the phase space distribution function is only a function of energy -- but is not a simple exponential. 

Our numerical simulations indicate that this spin segregation depends strongly on the difference in the trapping frequencies seen by the two spin species. Moreover, despite being in a nondegenerate regime, substantial quantum coherences are found in this system.  We believe that much can be learned from studying how these collisionless dynamics evolve into hydrodynamics as the scattering length is made larger.

\textit{Acknowledgements}:$-$~We thank John Thomas for stimulating discussions and Joseph Thywissen for useful comments.  S.N would like to thank Kaden R.A. Hazzard and Stefan K. Baur for useful conversations. This work was partially supported by NSF Grant No. PHY-0758104.


\begin{thebibliography}{99}

\bibitem{jt} X.Du, L.Luo, B.Clancy, and J.E. Thomas, Phys. Rev. Lett. \textbf{101} 150401 (2008).

\bibitem{johnson} B.R. Johnson, J.S. Denker, N.Bigelow, L.P.Levy, J.H. Freed, and D.M. Lee, Phys. Rev. Lett. \textbf{52} 1508 (1984).

\bibitem{levy} L.P. Levy and A.E. Ruckenstein, Phys. Rev. Lett. \textbf{52}, 1512 (1984).

\bibitem{oktel} M.\"{O}.Oktel and L.S. Levitov, Phys. Rev. Lett \textbf{88}, 230403 (2002).

\bibitem{clark} J.E.Williams, T. Nikuni, and Charles W. Clark, Phys. Rev. Lett. \textbf{88}, 230405 (2002).

\bibitem{laloe2}J.N.Fuchs, D.M.Gangardt, F. Lalo\"{e}, Phys. Rev. Lett. \textbf{88}, 230405 (2002) . 

\bibitem{lew} H.J. Lewandowski, D.M. Harber, D.L.Whitaker, and E.A. Cornell,  Phys. Rev. Lett. \textbf{88} 070403 (2002).

\bibitem{jtnew} X.Du, Y.Zhang, J.Petricka, and J.E.Thomas, \textit{www.arxiv.org/cond-mat/0901.3702v1}.

\bibitem{laloe} F.Pi\'{e}chon, J.N.Fuchs and F.Lalo\"{e}, \textit{www.arxiv.org/cond-mat/0901.4008v1}.

\bibitem{bigelow} N.P.Bigelow, J.H. Freed and D.M. Lee, Phys Rev. Lett., \textbf{63} 1609-1612 (1989). 

\bibitem{silin} V.P.Silin, Sov. Phys. JETP \textbf{6}, 945 (1958).

\bibitem{leggett} A.J.Legett, J.Phys C \textbf{3} 448 (1970).

\bibitem{laloeorig}  C. Lhuillier and F. Lalo\"{e}, J.Phys (Paris) \textbf{43} 197 (1982).

\bibitem{laloeorig2}  C. Lhuillier and F. Lalo\"{e}, J.Phys (Paris) \textbf{43} 225 (1982).

\bibitem{jeon} J. W. Jeon and W. J. Mullin, J. Phys. France \textbf{49} (1988).

\bibitem{kadbaym} \textit{Quantum Statistical Mechanics}, L. P. Kadanoff, and G. Baym, W. A. Benjamin, Inc. 1962. 

\bibitem{teuk} \textit{Numerical Recipes Third Edition: The Art of Scientific Computing} W.H.Press, S.A. Teukolsky, W.T.Vetterling, and B.P.Flannery, Cambridge University Press, 2007.

\bibitem{pethicksmith} \textit{Bose-Einstein Condensation in Dilute Gases} C.J.Pethick and H.Smith, Cambridge University Press, 2002.



\end{thebibliography}
\end{document}